\documentclass[12pt]{article}
\usepackage[utf8]{inputenc}
\usepackage{bbold}
\usepackage[title]{appendix}
\usepackage{comment}
\usepackage{float}
\usepackage{array}
\usepackage[]{natbib}
\usepackage{graphicx}
\usepackage{xcolor}

\usepackage[margin=1in]{geometry}
\usepackage{amsmath,amssymb}
\usepackage{booktabs}
\usepackage{url}
\usepackage[hidelinks]{hyperref}

\newcommand{\newc}{\newcommand}
\newc{\bc}{\begin{center}}
\newc{\ec}{\end{center}}
\newc{\be}{\begin{enumerate}}
\newc{\ee}{\end{enumerate}}
\newc{\bi}{\begin{itemize}}
\newc{\ei}{\end{itemize}}
\newc{\bd}{\begin{description}}
\newc{\ed}{\end{description}}
\newc{\und}[1]{\underline{#1}}
\newc{\E}{\mbox{E}}
\newc{\V}{\mbox{V}}
\newc{\N}{\mbox{N}}
\newc{\B}{\mbox{Bin}}
\newc{\Bern}{\mbox{Bern}}
\newc{\Po}{\mbox{Po}}
\newc{\IG}{\mbox{IG}}
\newc{\Gam}{\mbox{Gam}}
\newc{\bdP}{\mbox{P}}
\newc{\bdp}{\mathsf{p}}
\newc{\bdphat}{\hat{\mathsf{p}}}
\newc{\odds}{\mbox{odds}}
\newc{\OR}{\mbox{OR}}
\newc{\stderr}{\mbox{s.e.}}
\newc{\logit}{\mbox{logit}}
\newc{\sign}{\mbox{sign}}
\newc{\SD}{\mbox{SD}}
\newc{\bdmu}{\mbox{\boldmath $\mu$}}
\newc{\bdSigma}{\mbox{\boldmath $\Sigma$}}
\newc{\bdLambda}{\mbox{\boldmath $\Lambda$}}
\newc{\bdmuhat}{\mbox{\boldmath $\hat{\mu}$}}
\newc{\bdeta}{\mbox{\boldmath $\eta$}}
\newc{\bdtheta}{\mbox{\boldmath $\theta$}}
\newc{\bdbeta}{\mbox{\boldmath $\beta$}}
\newc{\bdgamma}{\mbox{\boldmath $\gamma$}}
\newc{\bdetahat}{\mbox{\boldmath $\hat{\eta}$}}
\newc{\bdbetahat}{\mbox{\boldmath $\hat{\beta}$}}
\newc{\bdgammahat}{\mbox{\boldmath $\hat{\gamma}$}}
\newc{\bdthetahat}{\mbox{\boldmath $\hat{\theta}$}}
\newc{\bdvareps}{\mbox{\boldmath $\varepsilon$}}
\newc{\bdzero}{\mbox{\boldmath $0$}}
\newc{\bdone}{\mbox{\boldmath $1$}}
\newc{\bdnu}{\mbox{\boldmath $\nu$}}
\newc{\bdell}{\mbox{\boldmath $\ell$}}
\newc{\bdxi}{\mbox{\boldmath $\xi$}}
\newc{\bdomega}{\mbox{\boldmath $\omega$}}
\newc{\bdepsilon}{\mbox{\boldmath $\varepsilon$}}
\newc{\bdI}{\mathbf{I}}
\newc{\bdX}{\mbox{\boldmath $X$}}
\newc{\bdA}{\mbox{\boldmath $A$}}
\newc{\bdB}{\mbox{\boldmath $B$}}
\newc{\bdC}{\mbox{\boldmath $C$}}
\newc{\bdD}{\mbox{\boldmath $D$}}
\newc{\bdG}{\mbox{\boldmath $G$}}
\newc{\bdJ}{\mbox{\boldmath $J$}}
\newc{\Ktil}{\tilde{K}}
\newc{\Khat}{\hat{K}}
\newc{\bda}{\mbox{\boldmath $a$}}
\newc{\bdb}{\mbox{\boldmath $b$}}
\newc{\bdc}{\mbox{\boldmath $c$}}
\newc{\bde}{\mbox{\boldmath $e$}}
\newc{\bdu}{\mbox{\boldmath $u$}}
\newc{\bdv}{\mbox{\boldmath $v$}}
\newc{\bdx}{\mbox{\boldmath $x$}}
\newc{\bdy}{\mbox{\boldmath $y$}}
\newc{\bdz}{\mbox{\boldmath $z$}}
\newc{\bdr}{\mbox{\boldmath $r$}}
\newc{\bdQ}{\mbox{\boldmath $Q$}}
\newc{\bdR}{\mbox{\boldmath $R$}}
\newc{\bdY}{\mbox{\boldmath $Y$}}
\newc{\bdT}{\mbox{\boldmath $T$}}
\newc{\bdW}{\mbox{\boldmath $W$}}
\newc{\bdWtil}{\tilde{\mbox{\boldmath $W$}}}
\newc{\bdH}{\mbox{\boldmath $H$}}
\newc{\bdL}{\mbox{\boldmath $L$}}
\newc{\bdU}{\mbox{\boldmath $U$}}
\newc{\bdV}{\mbox{\boldmath $V$}}
\newc{\Multinom}{\mbox{Multinom}}
\newc{\Var}{\mbox{Var}}
\newc{\var}{\mbox{var}}
\newc{\diag}{\mbox{diag}}
\newc{\thetahat}{\hat{\theta}}
\newc{\tr}{\mbox{tr}}
\newc{\phat}{\hat{p}}
\newc{\Xbar}{\bar{X}}
\newc{\xbar}{\bar{x}}
\newc{\Ybar}{\bar{Y}}
\newc{\ybar}{\bar{y}}
\newc{\dbar}{\bar{d}}
\newc{\yhat}{\hat{y}}
\newc{\bdyhat}{\mbox{\boldmath $\hat{y}$}}
\newc{\ytil}{\tilde{y}}
\newc{\ftil}{\tilde{f}}
\newc{\Ho}{\mbox{\bf H}_o}
\newc{\Ha}{\mbox{\bf H}_a}
\newc{\phatYX}{\phat_Y - \phat_X}
\newc{\SSG}{\mbox{SSG}}
\newc{\SSB}{\mbox{SSB}}
\newc{\SSE}{\mbox{SSE}}
\newc{\SST}{\mbox{SST}}
\newc{\SSR}{\mbox{SSR}}
\newc{\SSAB}{\mbox{SSAB}}
\newc{\MSG}{\mbox{MSG}}
\newc{\MSB}{\mbox{MSB}}
\newc{\MSE}{\mbox{MSE}}
\newc{\MST}{\mbox{MST}}
\newc{\MSAB}{\mbox{MSAB}}
\newc{\dfE}{\mbox{dfE}}
\newc{\dfG}{\mbox{dfG}}
\newc{\dfB}{\mbox{dfB}}
\newc{\dfT}{\mbox{dfT}}
\newc{\dfAB}{\mbox{dfAB}}
\newc{\muhat}{\hat{\mu}}
\newc{\betahat}{\hat{\beta}}
\newc{\alphahat}{\hat{\alpha}}
\newc{\etahat}{\hat{\eta}}
\newc{\phihat}{\hat{\phi}}
\newc{\sigmahat}{\hat{\sigma}}
\newc{\cl}{\centerline}
\newc{\R}{\mathbb{R}}
\newc{\trans}{^\mathsf{T}}
\newc{\xtx}{\bdX\trans\bdX}
\newc{\xxtxx}{\bdX(\xtx)^{-1}\bdX\trans}
\newc{\argmin}{\operatornamewithlimits{argmin}}
\newc{\argmax}{\operatornamewithlimits{argmax}}
\newc{\incg}[2]{
\includegraphics[width=#1\textwidth]{#2}
}

\title{Regularization in Paired Comparison Models via Pseudo-Games and Phantom Players}
\author{Mark E. Glickman\thanks{Address for correspondence:
Department of Statistics, Harvard University, 33 Oxford Street,
Cambridge, MA  02138.  Email address: {\tt glickman@g.harvard.edu}} \\
Harvard University}
\date{}

\begin{document}
\maketitle

\begin{abstract}
Paired comparison models are useful for estimating latent abilities or preferences from binary outcomes, but maximum likelihood estimation can be unstable or fail when the comparison graph is disconnected or nearly separated. Ridge regularization addresses these difficulties by shrinking ability parameters toward a common center, but it can obscure the simple likelihood interpretation that makes Bradley-Terry and Thurstone-Mosteller models attractive to practitioners. This paper describes two data-augmentation perspectives on regularization. The first adds fractional pseudo-games between every pair of competitors. The second adds a fixed-strength phantom player and gives each real competitor a weighted pseudo-win and pseudo-loss against that player. Both approaches yield finite, shrunken estimates; the phantom-player construction also resolves the usual location nonidentifiability without an explicit linear constraint. For the Bradley-Terry model, the two augmentations lead to transparent penalty functions that can be compared directly with ridge penalties. An application to the 2025 Major League Baseball regular season illustrates that tuned pseudo-game and phantom-player regularization can closely reproduce ridge-regularized strength estimates while retaining an intuitive augmented-data representation.
\end{abstract}


\addtolength{\baselineskip}{12pt}

\newpage

\section{Introduction}

Paired comparison data arise whenever items, teams, players, candidates, 
or products are compared two at a time.  
A standard goal is to infer latent merits or strengths from observed preference outcomes.
A common paired comparison modeling framework is the linear paired comparison model
\citep{David1988};
if item $i$ is compared with item $j$, the probability that $i$ is preferred to $j$ is
\begin{equation}
  p_{ij} = \Pr(Y_{ij}=1) = F(\theta_i-\theta_j),
  \label{eq:linear-pc}
\end{equation}
where $F$ is monotonically increasing and satisfies $F(x) = 1-F(-x)$.
Two frequent choices are
$F(x)=\{1+\exp(-x)\}^{-1}$ for the Bradley-Terry model 
\citep{BradleyTerry1952}, 
and $F(x)=\Phi(x)$ (the standard normal cumulative distribution function)
for the Thurstone-Mosteller model \citep{Thurstone1927,Mosteller1951}.  
The parameters $\theta_1,\ldots,\theta_J$, with $J$ the number of items being compared,
represent latent abilities or merits.  
Because the model depends only on differences in ability, an ordinary maximum
likelihood analysis typically imposes a linear constraint such as
$\sum_{j=1}^J \theta_j=0$.
In what follows, we use the language of wins and losses, although the same ideas apply 
to preference and choice data.

Despite the simplicity of (\ref{eq:linear-pc}), ordinary maximum likelihood estimation 
can be problematic.  
Ford's strong-connectivity condition gives a classical characterization of when finite
Bradley-Terry maximum likelihood estimates exist \citep{Ford1957}, 
with analogous existence conditions for other linear paired comparison models.
Informally, every nonempty set of competitors must both defeat and be defeated by 
competitors outside the set.  
When a dominant competitor never loses, a weak competitor never wins, one group always defeats 
another group, or disconnected groups never play one another, the likelihood can 
be maximized only on the boundary, yielding infinite ability differences.  
These failures are not merely hypothetical.  
They arise in sports with incomplete schedules, preference studies with sparse 
comparison designs, and online ranking systems in which new items enter continually.

A two-player example illustrates the issue.  
Suppose Player~A defeats Player~B in five consecutive games.  
If $p$ denotes Player~A's probability of defeating Player~B, the binomial likelihood is 
proportional to $p^5$, so the ordinary maximum likelihood estimate is $\hat p=1$.  
Under the Bradley-Terry or Thurstone-Mosteller models,
this corresponds to $\hat\theta_A-\hat\theta_B=\infty$; 
with the usual centering constraint, the individual estimates are $+\infty$ and $-\infty$.  
The infinite estimate is a mathematically faithful summary of the likelihood, 
but it is rarely a satisfactory prediction.  
A future match might still be won by Player~B, and five games are not enough to justify 
a literal probability of one.  

Regularization provides a principled remedy by allowing the data to indicate that
Player~A is stronger while preventing a finite sample from producing an infinite
ability gap.
Ridge penalization, one particular approach to regularization, replaces the 
log-likelihood $\ell(\boldsymbol\theta)$ by
\begin{equation}
  \ell_R(\boldsymbol\theta;\lambda)
  = \ell(\boldsymbol\theta) - \lambda \sum_{j=1}^J \theta_j^2,
  \qquad \lambda \geq 0,
  \label{eq:ridge}
\end{equation}
shrinking strengths toward zero and producing finite estimates even in separated 
or disconnected cases. 
Ridge regularization is foundational in statistical modeling \citep{Hastie2020}. 
For paired comparisons, \citet{VarinFirth2024} developed ridge-penalized Bradley-Terry 
and Thurstone-Mosteller estimation and proposed a tractable empirical-Bayes/composite-likelihood 
method for tuning the ridge penalty, making ridge estimation a natural benchmark here. 
The penalty in (\ref{eq:ridge}) also resolves the usual location nonidentifiability, 
because among equivalent additive 
shifts in the $\theta_j$, it selects the centered vector with 
smallest squared norm. 

This paper develops an alternative but closely related perspective: regularization by
adding pseudo-observations.
The first approach adds fractional pseudo-games to the observed comparison table, so that
each pair of competitors is treated as if it had a small number of additional balanced
results.
The second approach adds a known-strength phantom player and gives each real
competitor a weighted pseudo-win and pseudo-loss against that player.
Both constructions shrink estimated abilities, avoid infinite estimates in separated
or sparse comparisons, and preserve a familiar likelihood form.
They can also be implemented using standard logistic or probit regression routines,
which makes them attractive for applied analysts who already use generalized linear
model software.
The two constructions regularize different aspects of the model:
pseudo-games act on pairwise ability differences, whereas phantom-player comparisons
anchor individual abilities to a fixed reference value.
This distinction affects both identifiability and the form of the induced penalty, as
developed below.
The main contribution of this paper is to show that these two simple augmented-data
constructions yield interpretable regularization penalties for paired comparison models,
with tuning parameters that can be understood on the scale of additional comparison
information.

This pseudo-observation view connects the proposed methods to several related approaches.
In sparse binomial and contingency-table problems, adding a small constant to cells of
a $2\times 2$ table, often associated with the Haldane-Anscombe correction, prevents
undefined log odds ratios when cells are zero \citep{Haldane1956,Anscombe1956}.
Modern bias-reduction methods provide a more principled analog: Firth's adjustment
can be interpreted as a Jeffreys-prior penalty and, in some settings, as iteratively
adjusted pseudo-data \citep{Firth1993,KosmidisFirth2021}.
For Bradley-Terry-type models, penalized likelihood, bias reduction, singular
perturbation, and Bayesian approaches have also been used to stabilize estimation
\citep{Firth2005,TurnerFirth2012,Yan2016,CaronDoucet2012,PhelanWhelan2018}.
The contribution in the present paper is not simply to add generic binomial pseudo-counts.
Rather, the pseudo-observations are chosen to regularize the comparison graph itself:
either by inserting balanced pseudo-results between all pairs or by comparing every
competitor to a common reference player.

The remainder of the paper is organized as follows.  
Section~\ref{sec:methods} develops pseudo-game and phantom-player regularization, 
derives the associated penalized log-likelihoods, compares the induced penalties to 
ridge regularization, and discusses tuning and inference.  
Section~\ref{sec:mlb} applies the methods to the 2025 Major League Baseball regular season.  
Section~\ref{sec:conclusion} concludes with practical recommendations, 
connections to Bayesian modeling, limitations, and possible extensions.

\section{Pseudo-Games and Phantom Players}
\label{sec:methods}

For a tournament or league involving $J$ competitors, 
let $y_{ij}$ denote the number of times competitor $i$ defeats 
competitor $j$, with $y_{ii}=0$.  
For each unordered pair $i<j$, write $m_{ij}=y_{ij}+y_{ji}$ and $p_{ij}=F(\theta_i-\theta_j)$.  
Because $F(x)=1-F(-x)$, the probability that competitor $j$ defeats competitor $i$ is
$1-p_{ij}$.
Ignoring constants that do not depend on $\boldsymbol\theta$, the log-likelihood is
\begin{equation}
  \ell(\boldsymbol\theta)
  = \sum_{i<j}\{ y_{ij}\log p_{ij} + y_{ji}\log(1-p_{ij})\}.
  \label{eq:loglik}
\end{equation}

\subsection{Pseudo-game regularization}
\label{subsec:pseudogames}

The pseudo-game approach adds a small value $\delta$ to each of the two win counts
associated with every unordered pair of competitors. 
That is, in the augmented data set,
each competitor is credited with $\delta$ fractional wins and $\delta$ fractional 
losses against every other competitor.  
The augmented log-likelihood is
\begin{align}
  \ell_{\delta}(\boldsymbol\theta)
  &= \sum_{i<j}\{(y_{ij}+\delta)\log p_{ij} + (y_{ji}+\delta)\log(1-p_{ij})\} \\
  &= \ell(\boldsymbol\theta) + \delta \sum_{i<j}\log\{p_{ij}(1-p_{ij})\},
  \qquad \delta\geq 0.
  \label{eq:pseudogame}
\end{align}
The second term is a penalty because $0<p_{ij}(1-p_{ij})\leq 1/4$.  
It is maximized when $p_{ij}=1/2$ for all pairs, which occurs when 
all ability parameters are equal.  
Thus increasing $\delta$ shrinks ability differences toward zero.

This representation also resembles Bayesian power-prior methods, in which a
historical-data likelihood is raised to a power to control the amount of information
borrowed from that source \citep{IbrahimChen2000,IbrahimChenSinha2003}.  
The pseudo-games used here are not historical observations, but they similarly add a
weighted likelihood component whose influence is governed by a tuning parameter.
The key distinction is that the additional likelihood contribution is deliberately
constructed to stabilize the comparison graph rather than obtained from previous
studies.

The tuning parameter $\delta$ can be calibrated through a simple 
predictive question.
Suppose Player~A defeats Player~B in a single observed game, and suppose 
no other information about either player is available.
How large should the estimated probability be that Player~A defeats Player~B 
in a future game?
If the analyst wants this one-win/no-loss record to correspond to a 
pre-specified probability $q>1/2$, then the pseudo-game construction 
implies 
\begin{equation}
  q = \frac{1+\delta}{1+2\delta}.
  \label{eq:delta-calibration}
\end{equation}
Solving for $\delta$ gives
\[
\delta = \frac{1-q}{2q-1}.
\]
Thus, choosing $q$ close to one corresponds to a small amount of regularization: 
for example, $q=0.99$ gives $\delta=1/98$.
Choosing a less extreme value of $q$ corresponds to a larger $\delta$, 
because the single observed win is then interpreted as weaker evidence that 
Player~A will defeat Player~B again.

The pseudo-game penalty depends only on ability differences.  
Therefore it does not resolve the location nonidentifiability 
of (\ref{eq:loglik}); a linear constraint such as $\sum_j\theta_j=0$ is 
still necessary.  
Under the Bradley-Terry logit link, however, the connection to ridge regularization 
can be understood from the following relationship.
For $d_{ij}=\theta_i-\theta_j$,
\begin{equation}
  \log\{p_{ij}(1-p_{ij})\}
  = d_{ij} - 2\log(1+\exp d_{ij})
  = -2\log 2 - \frac{d_{ij}^2}{4} + O(d_{ij}^4)
  \label{eq:bt-pair-penalty}
\end{equation}
near $d_{ij}=0$,
which results from a Taylor series expansion of
$d_{ij} - 2\log(1+\exp d_{ij})$ around $d_{ij}=0$.

The aggregate quadratic term in this local expansion is proportional to the usual
ridge penalty under the centering constraint.  
To see this,
\begin{align}
  \sum_{i<j} d_{ij}^2 =
  \sum_{i<j}(\theta_i-\theta_j)^2
  &= \frac{1}{2}\sum_{i=1}^J\sum_{j=1}^J(\theta_i-\theta_j)^2 \notag \\
  &= \frac{1}{2}\sum_{i=1}^J\sum_{j=1}^J
  \left(\theta_i^2+\theta_j^2-2\theta_i\theta_j\right) \notag \\
  &= \frac{1}{2}
  \left\{
  J\sum_{i=1}^J\theta_i^2
  +
  J\sum_{j=1}^J\theta_j^2
  -
  2\sum_{i=1}^J\sum_{j=1}^J\theta_i\theta_j
  \right\} \notag \\
  &= J\sum_{j=1}^J\theta_j^2
  -
  \left(\sum_{j=1}^J\theta_j\right)^2.
  \label{eq:pairwise-diff-identity}
\end{align}
The first equality holds because the double sum over ordered pairs counts each 
unordered pair twice, while the terms with $i=j$ contribute zero.  
Therefore, under the usual centering constraint $\sum_{j=1}^J\theta_j=0$,
\begin{equation}
  \sum_{i<j} d_{ij}^2 =
  \sum_{i<j}(\theta_i-\theta_j)^2
  =
  J\sum_{j=1}^J\theta_j^2.
\end{equation}
Thus, in combination with~(\ref{eq:bt-pair-penalty}), 
pseudo-game regularization behaves locally like ridge regularization 
with approximate ridge coefficient $\lambda \approx \delta J/4$, up to an 
additive constant.

\subsection{Phantom-player regularization}
\label{subsec:phantom}

The phantom-player approach introduces an artificial competitor, indexed by $0$, 
whose ability is known and fixed at $\theta_0=0$.  
Each real competitor receives one pseudo-win and one pseudo-loss against the 
phantom player.  
These pseudo-comparisons are assigned weight $\rho\geq 0$ in the likelihood.  
Equivalently, when $\rho$ is an integer, the phantom-player contribution is the same
for likelihood maximization as adding $\rho$ wins and $\rho$ losses for each real
competitor against the fixed-strength phantom player; noninteger $\rho$ gives the
corresponding fractional weighted version.
The augmented log-likelihood is
\begin{align}
  \ell_{\rho}(\boldsymbol\theta)
  &= \ell(\boldsymbol\theta)
  + \rho \sum_{j=1}^J \left[\log F(\theta_j) + \log\{1-F(\theta_j)\}\right].
  \label{eq:phantom}
\end{align}
This penalty is maximized at $\theta_j=0$ for every $j$, so it shrinks abilities 
toward the known phantom-player strength.  
The phantom-player construction has the same discounted-likelihood flavor as
the pseudo-game approach, but the
additional comparisons are made against a common reference competitor rather than
between every pair of real competitors.
Unlike the pseudo-game approach, it acts directly on each $\theta_j$ rather than 
only on pairwise differences.  
Consequently, it resolves the location nonidentifiability without imposing a 
sum-to-zero constraint.

\begin{figure}[H]
\centering
\includegraphics[width=0.95\textwidth]{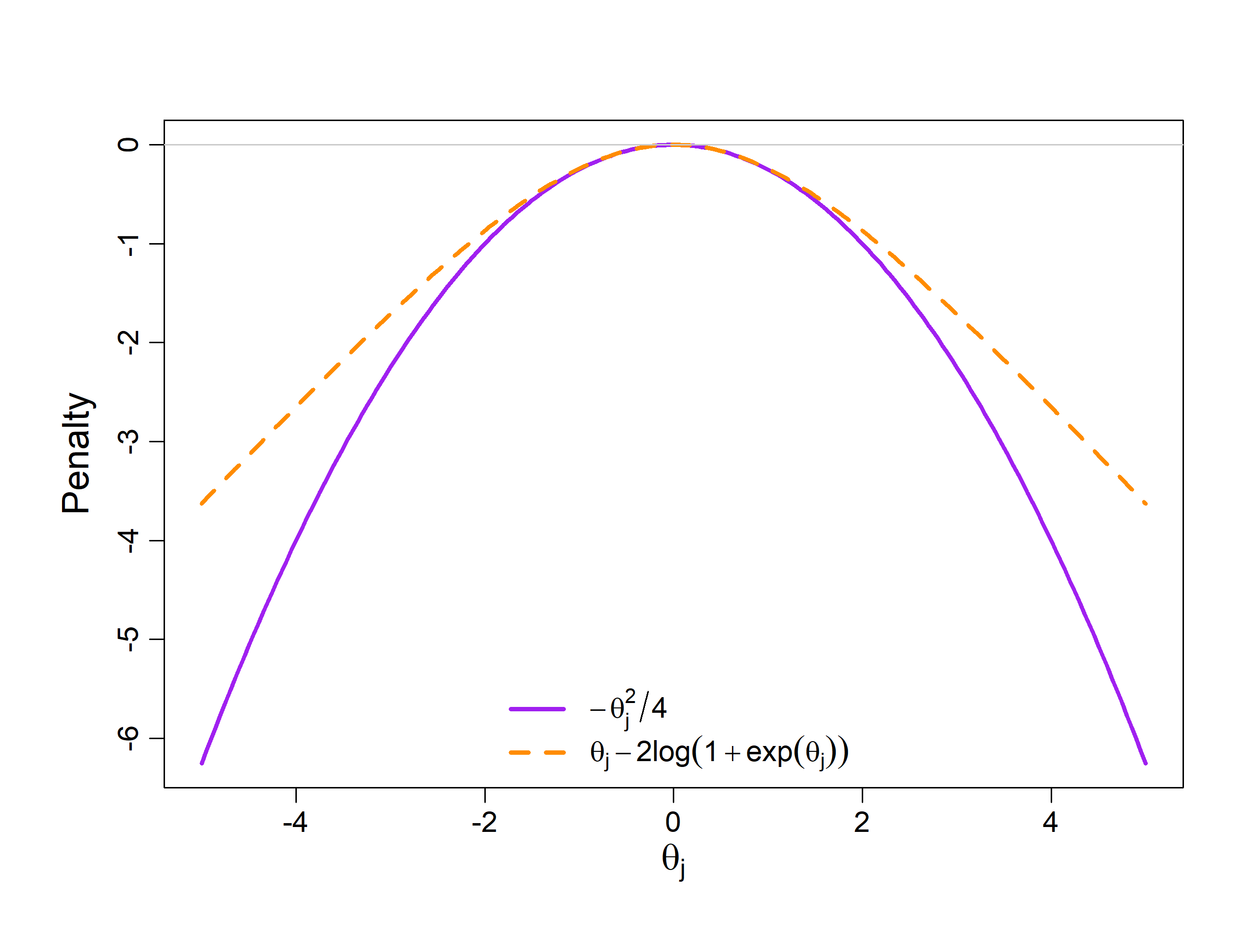}
\caption{Ridge penalty $-\theta^2/4$ and Bradley-Terry phantom-player penalty
$\theta-2\log(1+\exp\theta)$, shifted to have maximum zero.  The penalties have the
same curvature at $\theta_j=0$ after scaling, but the phantom-player penalty has approximately
linear rather than quadratic tails.}
\label{fig:penalties}
\end{figure}

For the Bradley-Terry logit link, (\ref{eq:phantom}) becomes
\begin{equation}
  \ell_{\rho}(\boldsymbol\theta)
  = \ell(\boldsymbol\theta)
  + \rho \sum_{j=1}^J \left[ \theta_j - 2\log(1+\exp\theta_j)\right].
  \label{eq:phantom-bt}
\end{equation}
A Taylor series expansion around $\theta_j=0$ gives
\begin{equation}
  \theta_j - 2\log(1+\exp\theta_j)
  = -2\log 2 - \frac{\theta_j^2}{4} + O(\theta_j^4),
  \label{eq:phantom-ridge}
\end{equation}
so phantom-player regularization is locally equivalent to ridge regularization 
with approximate ridge coefficient $\lambda\approx \rho/4$.  
The tails differ from a quadratic penalty, however.
For large positive or negative $\theta_j$, the logistic phantom penalty is 
approximately linear rather than quadratic. 
Figure~\ref{fig:penalties} displays this comparison.

\subsection{Implementation using standard regression software}
\label{subsec:implement}

Both regularization methods can be implemented using standard binomial regression software.
In the software package R \citep{RCoreTeam2025}, 
the most direct implementation uses the base function \texttt{glm} with a two-column
binomial response matrix. 
For the Bradley-Terry model, one uses
\texttt{family = binomial(link = "logit")}; for the Thurstone-Mosteller model, one uses
\texttt{family = binomial(link = "probit")}.

For pseudo-game regularization, create one row for each unordered pair $i<j$.
The response supplied to \texttt{glm} is the two-column matrix with row
\[
  (y_{ij}+\delta,\; y_{ji}+\delta)
\]
for pair $i<j$.
For noninteger $\delta$, R 
may issue the standard warning about noninteger binomial
counts; here the two-column response is being used as a convenient likelihood
specification with fractional case counts.
The corresponding design matrix consists of one row per comparison, and $J$ columns.
If row $r$ of the design matrix represents competitor $i$ evaluated against competitor
$j$, then row $r$ has a $1$ in column $i$, a $-1$ in column $j$, and zeros elsewhere,
so that the linear predictor is $\theta_i-\theta_j$.
Because this construction depends only on ability differences, the usual location
nonidentifiability remains. 
In a \texttt{glm} implementation, this can be handled by
dropping one competitor column and then recentering the fitted abilities, or by imposing
a sum-to-zero constraint through an equivalent reparameterization.
For example, one may construct a full $J$-column comparison matrix and then replace it 
by a $(J-1)$-column matrix that implicitly sets 
$\theta_J=-\sum_{j=1}^{J-1}\theta_j$, so that the fitted coefficient vector satisfies 
$\sum_{j=1}^J\theta_j=0$.  
This type of reparameterization is described in the team-strength modeling discussion of 
\citet{GlickmanStern2017}.

For phantom-player regularization, start with the ordinary paired comparison design
matrix and append $J$ additional rows.
These appended rows form the $J\times J$ identity matrix; row $j$ has linear predictor
$\theta_j$, corresponding to a comparison between real competitor $j$ and a known
phantom player whose strength is fixed at zero.
The response for each appended row is $(1,1)$, representing one pseudo-win and one
pseudo-loss against the phantom player.
The observed comparison rows receive case weight one, while the appended phantom-player
rows receive case weight $\rho$.
Thus the fit can be obtained with \texttt{glm} by passing the augmented two-column
response matrix, the augmented design matrix, and a \texttt{weights} vector with entries
equal to $1$ for real comparisons and $\rho$ for phantom-player comparisons.
This weighted binomial regression maximizes (\ref{eq:phantom-bt}).
Unlike the pseudo-game construction, the phantom-player construction anchors the
ability scale through the fixed-strength reference competitor, so no additional location
constraint is required. 
Estimates may still be centered after fitting for ease of
presentation, though this is not strictly necessary.

\subsection{Choosing tuning parameters}
\label{subsec:tuning}

The regularization parameters $\delta$ and $\rho$ may be set using 
subject-matter judgments or chosen by cross-validation.  
Expert elicitation is straightforward for $\delta$ through (\ref{eq:delta-calibration}), 
and for $\rho$ through the interpretation that each competitor receives 
$\rho$ weighted pseudo-wins and $\rho$ weighted pseudo-losses against a 
fixed-strength reference competitor.

For data-adaptive tuning, one can use $K$-fold cross-validation.  
For each candidate value of the tuning parameter, fit the regularized model 
on $K-1$ folds and evaluate the ordinary, unregularized log-likelihood on 
the held-out fold.  
The selected tuning value maximizes the summed validation log-likelihood.  
The same procedure applies to ridge regularization, with candidate values 
of $\lambda$.

Two cautions are worth mentioning.  
First, very small or highly separated data sets may not contain enough 
validation information to choose a stable tuning parameter.  
Second, if the tuning parameter is chosen by cross-validation, conventional standard 
errors from a final augmented-data fit condition on the selected tuning value and do 
not reflect all sources of uncertainty.  
A bootstrap that repeats the full modeling 
procedure, including tuning-parameter selection, better reflects the sampling 
variation of the estimator actually used, although at greater computational cost 
\citep{Harrell2015}.

In practice, it is helpful to examine the validation curve rather than only its maximizer. 
A flat curve indicates that many tuning values have similar predictive performance, 
while a sharply peaked curve suggests that the data contain more information about 
the amount of regularization. 
The folds should match the prediction task: games may be held out in sports applications,
whereas more stringent designs might hold out all comparisons involving selected teams
or players when the goal is to assess prediction for sparsely observed or newly entering
competitors.

\subsection{Estimation and inference}

Conditional on a fixed tuning parameter, point estimation can proceed by 
maximum likelihood on the augmented data, as described in Section~\ref{subsec:implement}.  
For pseudo-games, the fitted model is an ordinary paired comparison likelihood 
with fractional counts and a centering constraint.  
For phantom players, the fitted model is a weighted paired comparison likelihood 
with additional rows against the phantom competitor.  
Wald intervals based on the observed information from the augmented likelihood 
can be useful as approximate summaries of curvature, but they should be described 
as conditional on the regularization choice.  
In applications where interval estimates are central, a parametric or nonparametric 
bootstrap that repeats the tuning step, or a fully Bayesian analysis with proper priors, 
may be preferable.

The Bayesian connection is direct.  
Ridge regularization corresponds to maximizing a posterior density under 
independent normal priors centered at zero. 
For common links such as logit and probit, the phantom-player penalty can also be
interpreted as a maximum {\em a posteriori} (MAP) 
estimator under independent proper shrinkage priors with densities
proportional to $[F(\theta_j)\{1-F(\theta_j)\}]^\rho$.
Similarly, pseudo-game regularization corresponds to the pairwise prior-like factor
\[
  \prod_{i<j} [F(\theta_i-\theta_j)\{1-F(\theta_i-\theta_j)\}]^\delta,
\]
together with an identifying constraint.
Thus pseudo-observation regularization can be viewed either as augmented likelihood 
or as a particular prior-like shrinkage device.

\section{Application to 2025 Major League Baseball}
\label{sec:mlb}

We illustrate the methods using the 2025 Major League Baseball regular season.
The season included 30 teams, with each team scheduled to play 162 games.
The regular season began with the Tokyo Series between the Los Angeles Dodgers and
Chicago Cubs on March 18--19, 2025, followed by the league-wide Opening Day schedule
on March 27, 2025.
The resulting data set consisted of all 2,430 regular-season games, with each game
represented as a binary outcome indicating whether the home team defeated the
visiting team.
Game-level data were obtained using the \texttt{baseballr} R package
\citep{PettiGilani2022}, which provides tools for acquiring baseball data from
sources including the MLB Stats API.
Our baseline analysis fits an ordinary Bradley-Terry model without a home-field 
parameter.  
The regularized analyses fit 
(i) ridge-penalized Bradley-Terry, 
(ii) pseudo-game Bradley-Terry, and 
(iii) phantom-player Bradley-Terry.  
Ridge-penalized fits were computed in R using the \texttt{glmnet} package
\citep{FriedmanHastieTibshirani2010}.
Because the purpose of the example is to compare regularization mechanisms, 
home-field advantage is omitted from the analyses; 
it could be added as an ordinary covariate in all three approaches.
The example is therefore intended as a comparison of regularization mechanisms rather
than as a definitive ranking model for MLB team quality.

Baseball is a useful illustration because the ordinary comparison graph is connected 
and the unregularized estimates exist, yet the schedule is still highly structured.  
Teams play division opponents more often than non-division opponents, so the schedule is 
not a balanced round robin in which every pair of teams competes equally often.
Regularization therefore has a role even when it is not strictly needed for existence: 
it controls the extent to which an extreme record against a particular schedule 
is translated into an extreme latent ability.  

\begin{figure}[H]
\centering
\includegraphics[width=0.82\textwidth]{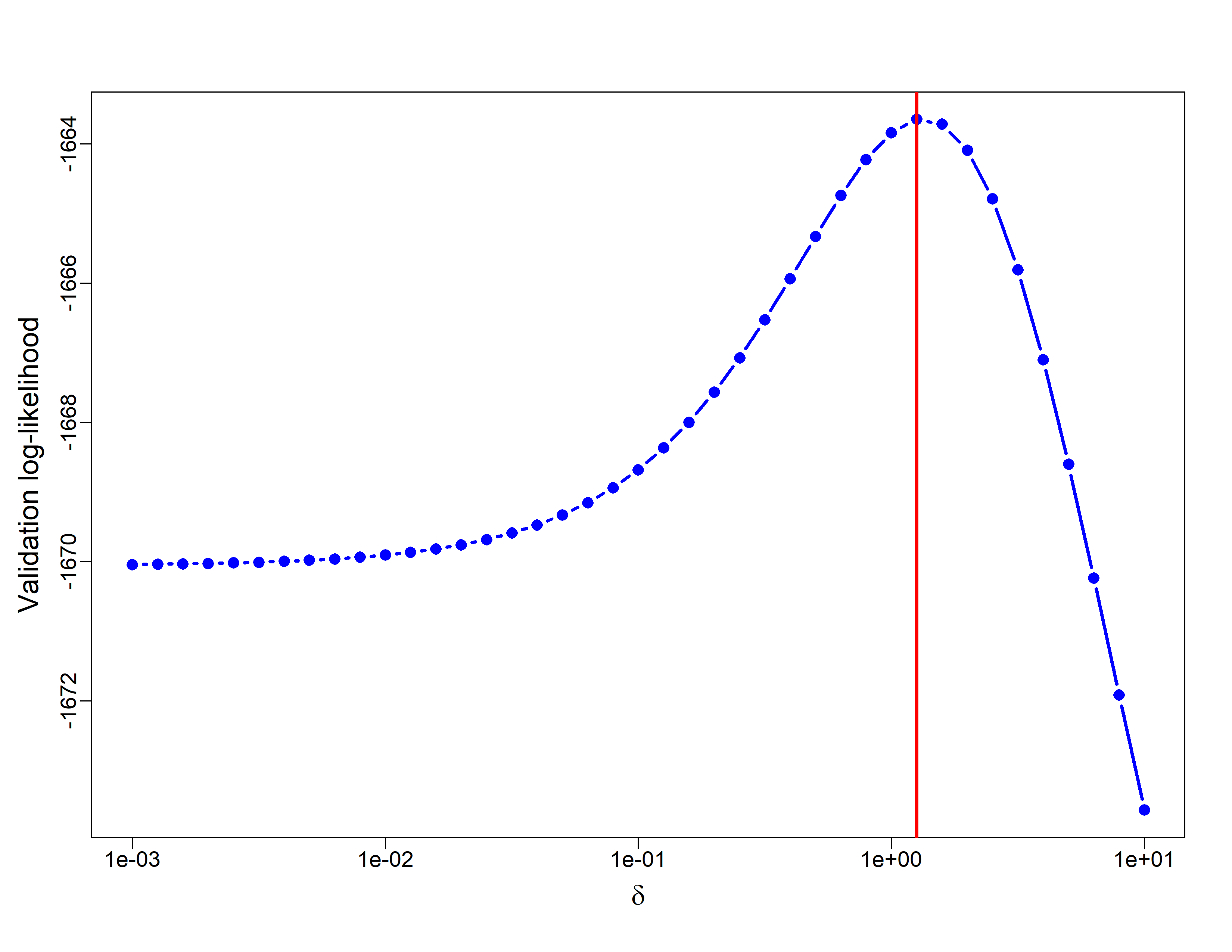}
\caption{Ten-fold validation log-likelihood for the pseudo-game tuning parameter $\delta$ in the MLB application. The vertical line marks the selected value, $\delta=1.2589$.}
\label{fig:deltaopt}
\end{figure}

\begin{figure}[H]
\centering
\includegraphics[width=0.82\textwidth]{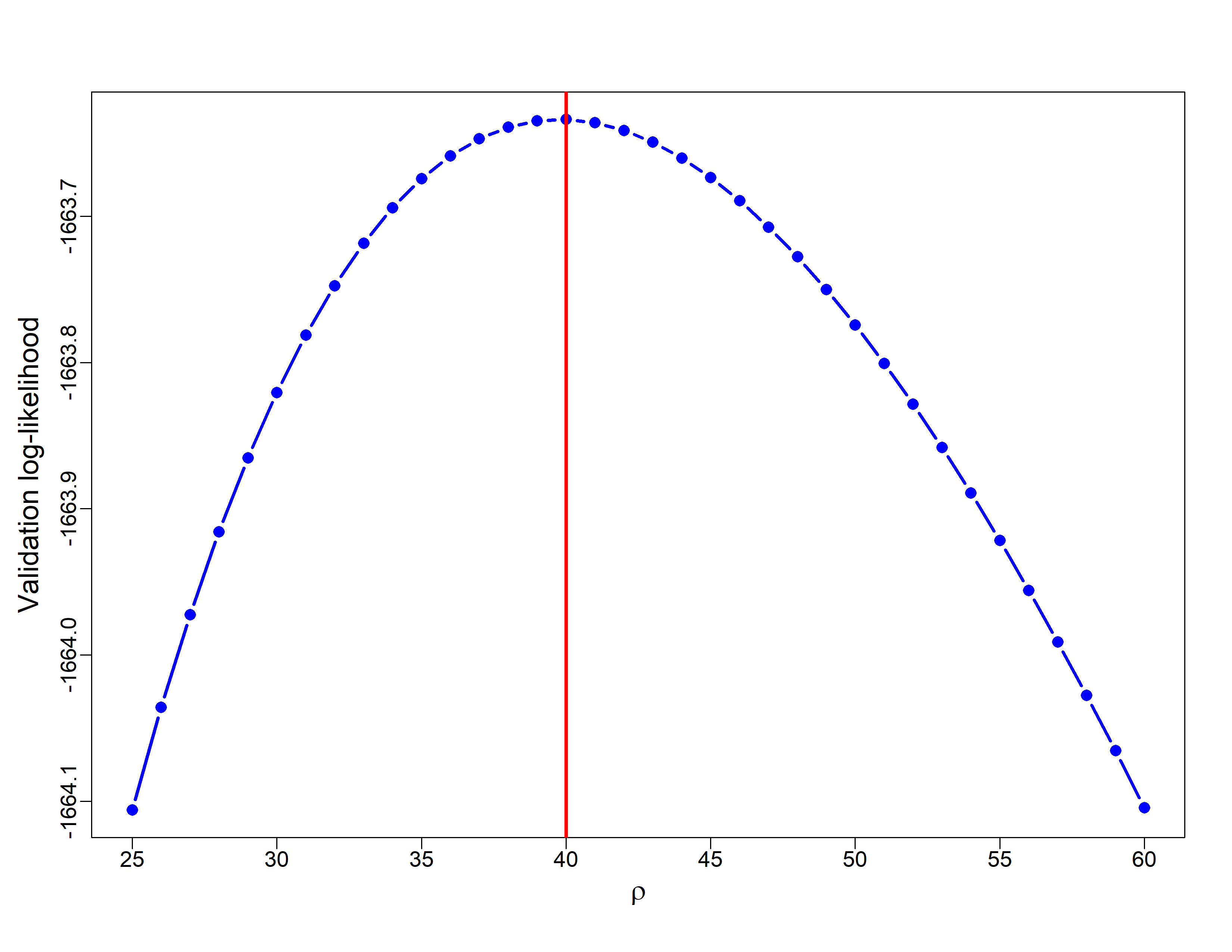}
\caption{Ten-fold validation log-likelihood for the phantom-player tuning parameter $\rho$ in the MLB application. The vertical line marks the selected value, $\rho=40$.}
\label{fig:rhoopt}
\end{figure}

The ordinary Bradley-Terry model is well-defined for these data, 
and was fit using the \texttt{glm} function in~R, imposing the sum-to-zero linear
constraint on the team strength parameters.
Under the \texttt{glmnet} objective  function, 
10-fold cross-validation for the ridge fit resulted in a tuning value of 0.01.
For the pseudo-game approach, candidate $\delta$ values ranged from 
$0.001$ to $10$ on a logarithmic grid, and 10-fold cross-validation 
selected $\delta=1.2589$ (Figure~\ref{fig:deltaopt}).  
For the phantom-player approach, candidate $\rho$ values ranged from $25$ to $60$, 
and 10-fold cross-validation selected $\rho=40$ (Figure~\ref{fig:rhoopt}).
For comparability, the same 10 folds were used for all three cross-validated
regularization methods.  
Folds were formed at the game level, so that each held-out
set consisted of observed games rather than held-out teams.  
Validation scores were
computed using the ordinary Bradley-Terry log-likelihood on the 
held-out games, without
pseudo-games, phantom-player rows, or ridge penalties.

\subsection{Comparison of fitted strengths}

Table~\ref{tab:mlb-selected} reports selected fitted strengths from the 
ordinary Bradley-Terry model and the three regularized fits.  
The table displays the top teams, middle teams near the league average, and bottom teams.  
The ordinary estimates show the largest spread.  
The pseudo-game and phantom-player estimates are substantially shrunken, 
and the cross-validated phantom-player estimates are especially close to the 
ridge estimates.

\begin{table}[ht]
\centering
\begin{tabular}{lrrrr}
\hline
Team & BT & Ridge & Pseudo-game & Phantom \\
\hline
Milwaukee Brewers      &  0.386 &  0.240 &  0.263 &  0.258 \\
Toronto Blue Jays      &  0.347 &  0.207 &  0.228 &  0.223 \\
Philadelphia Phillies  &  0.344 &  0.219 &  0.239 &  0.234 \\
New York Yankees       &  0.337 &  0.203 &  0.224 &  0.219 \\
Chicago Cubs           &  0.265 &  0.165 &  0.181 &  0.177 \\ \hline
Kansas City Royals     &  0.012 &  0.009 &  0.010 &  0.010 \\
Texas Rangers          &  0.010 &  0.003 &  0.004 &  0.004 \\
San Francisco Giants   & -0.023 & -0.008 & -0.010 & -0.009 \\ 
Los Angeles Angels     & -0.193 & -0.128 & -0.138 & -0.136 \\ \hline
Pittsburgh Pirates     & -0.220 & -0.143 & -0.155 & -0.152 \\
Minnesota Twins        & -0.262 & -0.165 & -0.180 & -0.177 \\
Washington Nationals   & -0.372 & -0.229 & -0.251 & -0.246 \\
Chicago White Sox      & -0.502 & -0.314 & -0.344 & -0.337 \\
Colorado Rockies       & -0.979 & -0.580 & -0.643 & -0.629 \\
\hline
\end{tabular}
\caption{Selected 2025 MLB strength estimates under ordinary and regularized
Bradley-Terry fits. The rows include leading teams, teams near the league average,
and lower-ranked teams. Values are on the log-ability scale.
The regularized columns use cross-validated tuning parameters.}
\label{tab:mlb-selected}
\end{table}

The shrinkage is easiest to see for the most extreme teams.  
The Colorado Rockies have an ordinary Bradley-Terry estimate of $-0.979$, but the 
ridge, pseudo-game, and phantom-player estimates are $-0.580$, $-0.643$, and $-0.629$, 
respectively.  
The Milwaukee Brewers have an ordinary Bradley-Terry estimate of $0.386$, 
while the regularized estimates fall between $0.240$ and $0.263$.  
Teams near the league average, such as the Royals and Rangers, change very 
little because their ordinary estimates are already close to zero.

\begin{figure}[H]
\centering
\includegraphics[width=0.95\textwidth]{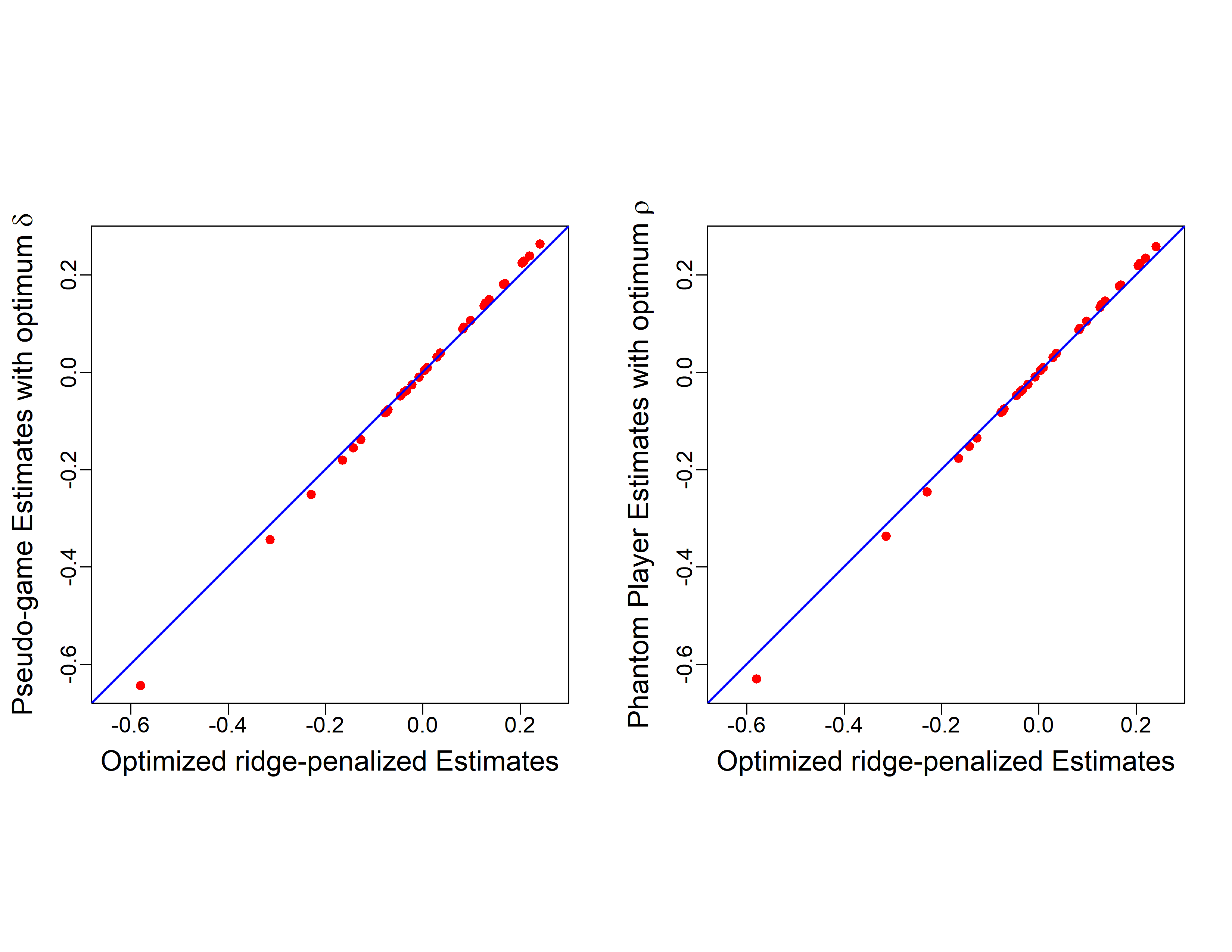}
\caption{Comparison of optimized ridge-penalized strength estimates with optimized 
pseudo-game estimates (left) and optimized phantom-player estimates (right) 
for the MLB application. 
The left panel uses $\delta=1.2589$ and the right panel uses $\rho=40$.
The blue line is the identity line.}
\label{fig:ridgecompare}
\end{figure}

The selected pseudo-game and phantom-player tuning values show how different 
augmented-data interpretations can lead to similar fitted strengths.  
The pseudo-game method with $\delta=1.2589$ adds more than one fractional win and 
one fractional loss to every pair of teams.  
The phantom-player method with $\rho=40$ adds one pseudo-win and one pseudo-loss 
against a fixed-strength reference opponent, each with case weight 40.  
Equivalently, for likelihood maximization, this is as if each team had been assigned 
40 wins and 40 losses, or 80 total games, against a team with strength fixed 
at zero.  
Figure~\ref{fig:ridgecompare} shows that both sets of estimates are close to the 
ridge estimates, with the phantom-player estimates especially close over the fitted range.

Table~\ref{tab:spread} summarizes the spread of selected fitted strengths.  
The ordinary Bradley-Terry fit has the widest range, mainly because of the 
very negative estimate for Colorado.  
All three regularized fits reduce the top-to-bottom range by roughly one 
third to two fifths.
The phantom-player range is closest to the ridge range, consistent with the 
point-by-point comparison in Table~\ref{tab:mlb-selected}.

\begin{table}[t]
\centering
\begin{tabular}{lrrr}
\hline
Model & Milwaukee & Colorado & Difference \\
\hline
Ordinary BT  & 0.386 & -0.979 & 1.365 \\
Ridge        & 0.240 & -0.580 & 0.820 \\
Pseudo-game  & 0.263 & -0.643 & 0.907 \\
Phantom      & 0.258 & -0.629 & 0.887 \\
\hline
\end{tabular}
\caption{Approximate spread of selected fitted strengths in the MLB example.
The difference is computed as Milwaukee's estimate minus Colorado's estimate
from Table~\ref{tab:mlb-selected}.}
\label{tab:spread}
\end{table}

The fitted strengths are on a log-odds scale.  
Consequently, changes that look modest in the table can have noticeable effects 
on predicted probabilities.  
A one-unit difference corresponds to an odds ratio of $\exp(1)\approx 2.72$ for 
a neutral comparison under the Bradley-Terry model.  
Shrinking the Milwaukee--Colorado difference from 1.365 to 0.887 under phantom-player 
regularization reduces the implied odds ratio from about 3.92 to about 2.43.

The regularized estimates identify the Brewers, Phillies, Blue Jays, Yankees, 
Cubs, Dodgers, Mariners, and Red Sox as among the strongest teams, 
with Milwaukee receiving the highest fitted strength in all models considered. 
At the lower end, the Rockies are a clear outlier, followed by the White Sox, 
Nationals, Twins, and Pirates. 
Regularization does not substantially alter these conclusions, but it 
reduces the implied separation between the best and worst teams.
For example, the ordinary Bradley-Terry difference between Milwaukee and Colorado 
is approximately $1.365$, corresponding to a neutral-field win probability of $0.797$. 
The phantom-player difference is approximately $0.887$, corresponding to a probability 
of $0.708$. 
Both probabilities favor Milwaukee strongly, but the regularized estimate is more
conservative and reflects the predictive shrinkage selected by cross-validation.

\section{Conclusions}
\label{sec:conclusion}

Pseudo-games and phantom players provide intuitive ways to regularize 
paired comparison models.  
Pseudo-game regularization adds fractional wins and losses to each pairwise matchup, 
thereby shrinking pairwise ability differences.  
Phantom-player regularization adds a fixed-strength reference opponent and gives each
competitor balanced pseudo-results against that opponent, thereby shrinking 
each ability directly toward the fixed reference strength.  
In the Bradley-Terry model, both approaches yield explicit penalty functions 
and are locally comparable to ridge regularization.

The main practical appeal is that the likelihood form remains familiar.  
Analysts can fit the resulting models using standard binomial regression routines 
with augmented responses, design matrices, and weights.  
In applications where the main concern is separation between observed pairs, pseudo-games
provide a direct pairwise correction; when the main concern is anchoring poorly connected
or disconnected competitors to a common scale, the phantom-player construction is often
simpler.

The approaches also have limitations.  
Pseudo-game regularization does not remove the location nonidentifiability of 
ordinary paired comparison models, so a centering constraint is still required.  
Both methods require a tuning parameter, and uncertainty summaries should account 
for whether that parameter was fixed in advance or selected by cross-validation.  
In sparse or highly separated examples, cross-validation may be unstable, and 
expert calibration or Bayesian modeling with a specified shrinkage prior may be 
more appropriate.

For routine use, analysts can fit the ordinary paired comparison model when it exists, 
check the comparison graph for disconnected or nearly separated components, and 
compare ridge, pseudo-game, and phantom-player fits over sensible tuning grids. 
Agreement across methods supports robust conclusions; disagreement may be a useful 
diagnostic and should prompt closer examination of the comparison graph, tuning choices, 
and influential observations.
The tuning procedure and the treatment of tuning uncertainty should be reported.

Several extensions are immediate.  
Covariates, such as home-field advantage, can be included without changing the 
regularization structure.  
While the development in this paper focused on the Bradley-Terry model,
the same ideas can be developed for Thurstone-Mosteller models by using the probit 
link in the augmented binomial regression.  
Another possible extension is to replace the single zero-strength phantom player by several
phantom opponents with fixed, dispersed abilities.  
Such a construction would induce a
penalty centered on a reference distribution of strengths rather than a 
single point, and
could provide more flexible tail behavior.
Extensions to ties, such as Davidson's model \citep{Davidson1970}, and to 
rank-order data, such as Plackett-Luce models \citep{Luce1959,Plackett1975}, 
are promising directions.  
For ties, pseudo-observations could be constructed to add small amounts of win, 
loss, and tie evidence, with separate consideration of the tie parameter.  
For rankings, phantom items or fractional pseudo-rankings could anchor strength
parameters and prevent degenerate rankings in sparse data.  
In both settings, the central question is the same as in the binary case: 
how should the augmented observations be interpreted, and what penalty do they induce?

More broadly, pseudo-observation regularization gives practitioners a bridge 
between penalized optimization and likelihood-based modeling.  
Regularization can be described not only as an abstract penalty, but also as carefully 
controlled additional information.  
This interpretation is especially useful in paired comparison problems, where the 
source of instability is often visible in the comparison graph: some competitors are 
poorly connected, some comparisons are too one-sided, or some groups of competitors 
are only weakly linked.  
Pseudo-games and phantom players address these features directly by adding balanced, 
interpretable comparison information.  
As a result, the amount of regularization can be discussed in the same language as the 
data themselves: fractional games, weighted wins and losses, or comparisons to a 
reference opponent, rather than only through the scale of a penalty parameter.

\addtolength{\baselineskip}{-12pt}

\bibliographystyle{agsm}
\bibliography{refs}

\addtolength{\baselineskip}{12pt}

\end{document}